\documentclass[%
 reprint,
groupedaddress,
 amsmath,amssymb,
 aps,
 longbibliography,
]{revtex4-1}

\providecommand\bnabla{\boldsymbol{\nabla}}
\providecommand\bkappa{\boldsymbol{\kappa}}

\newcommand{\vpar}{\ensuremath{v_{\parallel}}}
\newcommand{\vperp}{\ensuremath{v_{\perp}}}
\newcommand{\xpar}{\ensuremath{x_{\parallel}}}
\newcommand{\xperp}{\ensuremath{x_{\perp}}}

\def \al {\mbox{$\alpha$}}

\def \kpar {\mbox{$k_{\parallel}$}}
\def \kperp {\mbox{$k_{\perp}$}}
\def \Lpar {\mbox{$L_{\parallel}$}}
\def \vth {\mbox{$v_{\mathrm{T}}$}}

\def \l {\mbox{$\ell$}}

\def \eps {\mbox{$\epsilon$}}

\def \ost {\mbox{$\widetilde{\omega}_*$}}

\def \od {\mbox{$\omega_d$}}

\def \odt {\mbox{$\widetilde{\omega}_d$}}

\def \Reff {\mbox{$R_{\mathrm{eff}}$}}

\usepackage{graphicx}
\usepackage{dcolumn}
\usepackage{bm}
\usepackage{amsmath}
\usepackage{color}
\usepackage{mathtools}

\begin{document}

\title{Critical gradient turbulence optimization toward a compact stellarator reactor concept}

\author{G. T. Roberg-Clark}
\email{gar@ipp.mpg.de}
\author{G. G. Plunk}
\author{P. Xanthopoulos}
\author{C. Nührenberg}
\author{S. A. Henneberg}
\author{H. M. Smith}
\affiliation{Max-Planck-Institut f\"ur Plasmaphysik, D-17491, Greifswald, Germany}


\date{\today}

\begin{abstract}
Integrating turbulence into stellarator optimization is achieved by targeting the onset for the ion-temperature-gradient mode, highlighting effects of field line curvature, parallel connection length, local magnetic shear, and flux surface expansion. The result is two compact quasihelically symmetric stellarator configurations, one of which admits a set of modular coils, with significantly reduced turbulent heat fluxes compared to a known stellarator. This new configuration combines low values of neoclassical transport, good alpha particle confinement, and Mercier stability at a plasma beta of almost 2$\%$.
\end{abstract}

\date{\today}

\maketitle


\textit{Introduction.}--  A primary obstacle for the success of magnetic confinement fusion is the transport caused by instabilities such as the ion temperature gradient (ITG) mode, which is thought to significantly reduce plasma confinement in experiments such as the Wendelstein 7-X stellarator \citep{Beurskens2021a,Baehner2021,Carralero2021}. To overcome the losses from such turbulence, a given configuration can be scaled up in size and heating power.  A less costly alternative, currently explored, is to shape the magnetic field to alleviate the turbulence. This option could be particularly appealing for reactor scenarios, in which it will likely be difficult to achieve density gradient stabilization of turbulence via pellet injections \citep{Bozhenkov2020a,Xanthopoulos2020}, since the penetration distance of pellets may be limited in comparison to the minor radius of a reactor.

To achieve turbulence optimization via shaping, several strategies have been developed to reduce the rate that turbulent transport increases (``stiffness") as a function of the ion temperature gradient \citep{Mynick2010a,Xanthopoulos2014a,Hegna2018,Nunami2013,Hegna2022}. Another approach is to target the onset (``critical") gradient of significant turbulent transport \citep{Roberg-Clark2021,Roberg-Clark2022,Zocco2022}, which relies primarily on linear physics of ITG modes themselves \citep{Jenko2001,Biglari1989a,Baumgaertel2013,Zocco2018} and avoids the hard problem of solving turbulence in the full range of toroidal geometries. Here we demonstrate optimization using a critical gradient (CG) approach, which even leads to reduced stiffness of ITG turbulence in the nonlinear regime \citep{Roberg-Clark2022}, albeit with some implied trade-offs for integrated stellarator optimization.

In this Letter, we first show CG optimization targeting the absolute threshold for ITG modes, producing the largest critical gradient of all stellarators known to us, while sacrificing magneto-hydrodynamic (MHD) stability. We then show that without compromising MHD stability, or other key properties needed by a stellarator design, one may target the CG of only the toroidal branch of the ITG mode, based on the assumption that turbulence intensity will be small below this threshold. This model highlights the familiar stabilizing effects of local shear, but gives greater emphasis to short connection lengths between regions of ``good'' and ``bad'' magnetic curvature.  The resulting objective function is used via optimization to produce a quasi-helically-symmetric configuration with strongly reduced ITG turbulence compared to a known stellarator experiment (HSX) \citep{Talmadge2008}, in addition to acceptable levels of neoclassical losses, alpha particle confinement, coil complexity, and MHD stability, thus completing the picture for an initial stellarator concept with improved ion confinement.

\textit{Definitions.}--  Following \citep{Plunk2014a}, we use the standard gyrokinetic system of equations \citep{Brizard2007} to describe electrostatic fluctuations destabilized along a thin flux tube tracing a magnetic field line. The ballooning transform \citep{Dewar1983a,Connor1978} is used to separate out the fast perpendicular (to the magnetic field) scale from the slow parallel scale. The magnetic field representation in field following (Clebsch) representation reads, $\mathbf{B}=\bnabla \psi \times \bnabla \al$, where $\psi$ is a flux surface label and $\alpha$ labels the magnetic field line on the surface. The perpendicular wave vector is then expressed as $\mathbf{k_{\perp}} = k_{\alpha} \bnabla \alpha + k_{\psi} \bnabla \psi$, where $k_{\alpha}$ and $k_{\psi}$ are constants, so the variation of $\mathbf{k_{\perp}}(\ell)$ stems from that of the geometric quantities $\bnabla \alpha$ and $\bnabla \psi$, with $\l$ the field-line-following (arc length) coordinate.

We assume Boltzmann-distributed (adiabatic) electrons, thus solving for the perturbed ion distribution $g_{i}(\vpar,\vperp,\l,t)$, defined to be the non-adiabatic part of $\delta f_{i}$ ($\delta f_{i}=f_{i}-f_{i0})$ with $f_{i}$ the ion distribution function and $f_{i0}$ a Maxwellian. The electrostatic potential is $\phi(\mathbf{\l})$, and $\vpar$ and $\vperp$ are the particle velocities parallel and perpendicular to the magnetic field, respectively. The gyrokinetic equation reads
\begin{equation}
i\vpar \frac{\partial g}{\partial \ell} + (\omega - \odt)g = \varphi J_0(\omega - \ost^{T})f_0\label{gk-eqn}
\end{equation}
where $\omega$ is the mode frequency, $\ost^{T} = (Tk_{\alpha}/q)\mathrm{d}\ln T/\mathrm{d}\psi \left(v^2/\vth^2 - 3/2\right)$ is the diamagnetic frequency, and $J_{0} = J_{0}(k_{\perp}(\l)v_{\perp}/\Omega(\l))$ is the Bessel function of zeroth order. The thermal velocity is $\vth = \sqrt{2T/m}$, the thermal ion Larmor radius is $\rho = \vth/(\Omega\sqrt{2})$, $n$ and $T$ are the background ion density and temperature, $q$ is the ion charge, $\varphi = q\phi/T$ is the normalized electrostatic potential, and $\Omega=q B/m$ is the cyclotron frequency, with $B=|\mathbf{B}|$. The magnetic drift frequency in the low $\beta$ approximation is $\odt = (1/\Omega)(\mathbf{k_{\perp}} \cdot \mathbf{b} \times \bkappa) (\vpar^2 + \vperp^2/2)=\omega_{d}(\l)(\vpar^2 + \vperp^2/2)$, with $\bkappa = \mathbf{b} \cdot\bnabla\ \mathbf{b}$ and $\mathbf{b}=\mathbf{B}/B$. For simplicity in this analysis, we set $k_{\psi}=0$. We rewrite the drift frequency as $\od(\ell) \propto K_{d}(\ell) \equiv a^2{\bnabla}\alpha \cdot \mathbf{b} \times \bkappa$, referring to $K_{d}(\ell)$ as the ``drift curvature'' and to individual regions of bad curvature along the field line (where $K_{d}>0$) as ``drift wells''. We define a radial coordinate $r=a\sqrt{\psi/\psi_{edge}}$, with $a$ the minor radius corresponding to the flux surface at the edge, and $\psi_{edge}$ the toroidal flux at that location. The temperature gradient scale length is measured relative to the minor radius, $a/L_{T}=-(a/T)\mathrm{d}T/\mathrm{d}r$. To study the most unstable ITG mode conditions, we have neglected certain stabilizing factors such as the density gradient \citep{Proll2022,Jorge2021b} and plasma beta (electromagnetic effects) \citep{Pueschel2008a}.

Finally, the gyrokinetic system is completed by quasineutrality,
\begin{equation}
\int d^3{\bf v} J_{0} g = n(1 + \tau) \varphi\label{eqn:qn},
\end{equation}
with $\tau=|q_{e}|T/(qT_{e})$, $T_{e}$ the electron temperature, and $q_{e}$ the electron charge.

\textit{Thresholds for ITG modes.} As argued in \citep{Roberg-Clark2022}, the CG can be estimated by using the model 
\begin{equation}\label{eqn:abscritgrad}
 \frac{a}{L_{\text{T,crit,abs}}}=2.66\left(\frac{a}{\Reff} + 8.00 \frac{a}{\Lpar_{\text{Floquet}}}\right)   
\end{equation}
where $\Reff$ is the local effective radius of curvature determined by the profile of $K_{d}(\ell)$, and the effective parallel connection length for modes near the absolute threshold, $\Lpar_{\text{Floquet}}$ [see discussion above eqn. (4) in \citep{Roberg-Clark2022}], is determined by the relative size of good curvature outside the drift well, which may stabilize extended Floquet-like modes. The effect we seek to enhance is contained in the first term proportional to $a/\Reff$ and thus to the size of ``bad'' curvature on the outboard midplane.

\begin{figure}
    \centering
    \includegraphics[scale=1.9]{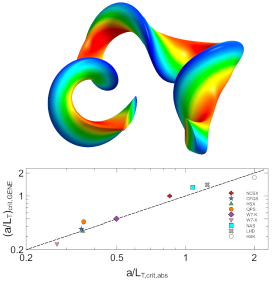}
    \caption{Above: the boundary surface of HSK showing the contours of B, with one half of a field period removed. Below: figure (1) of Roberg-Clark et al. (2022) plotting critical gradients found with GENE versus the model eqn. (\ref{eqn:abscritgrad}), with the point for HSK added.}
    \label{fig:HSK}
\end{figure}

It is expected, however, \citep{Jenko2002a,Plunk2017,Zocco2022} that the onset of toroidal ITG modes, as can be inferred from linear spectra in gyrokinetic simulations \citep{Zocco2018}, should lead to noticeable increases in nonlinear heat fluxes at a second, larger CG. The turbulence found below this onset (in the Floquet-like or slab-like regime) is thought to be more benign. We therefore also focus on the toroidal ITG mode \citep{Jenko2001,Plunk2014a,Zocco2018,Sugama1999a} with strongly peaked eigenmode structure that decays within a single drift well. In the local (in $\l$) theory of toroidal ITG modes \citep{Biglari1989a,Jenko2001,Plunk2014a}, the CG is set by the drive parameter $\kappa_{d}= \Reff/L_{T} $. This threshold can be computed for general parameter $b = \kperp^2\rho^2$  by solving the local dispersion relation 
\begin{equation} \label{eqn:torInt}
0=2-\frac{2}{\sqrt{\pi}}\int_{0}^{\infty}d \xperp \: \xperp \int_{-\infty}^{\infty}d \xpar \left[ \frac{\omega-\ost}{\omega - \odt} \right] J^{2}_{0} \exp(-x^2),
\end{equation}
which upon substitution of $K_d = a/\Reff$ yields $\Reff / L_{T,crit} = F(b)$, where $F(b)$ can be obtained numerically and is fairly well approximated by
 \begin{align}
     F(b)= 2.84+4.926 \: b, b < 0.755 \nonumber \\
           0.0371+7.51 \sqrt{b}, b \ge 0.755
 \end{align}
In realistic geometry, the threshold is controlled by the extent of drift wells, {\em i.e.} the parallel connection length \Lpar, but this can be related to finite Larmor radius (FLR) stabilization as follows: Note that a toroidal mode must have a drift frequency $\od \propto k_\alpha$ that exceeds the parallel transit rate $\kpar v_{T} \sim \pi v_{T}/\Lpar$.  Although this can always be satisfied by choice of $k_\alpha$, the increase of $k_\alpha$ comes at the cost of increasing $b$ as $\Lpar$ is reduced.  Thus, to estimate the critical gradient, we simply determine the minimum value of $b$ for which the resonance condition is satisfied, namely that for which $\od \sim \pi v_{T}/\Lpar$, yielding $b_{min} = (\pi a |\bnabla \alpha| \Reff/\Lpar)^{2}$, and

\begin{equation}\label{eqn:crit}
\frac{a}{L_{T,crit}}=\frac{a}{\Reff}F((\pi a|\bnabla \alpha| \Reff/\Lpar)^{2})
\end{equation}

\noindent with $F(b)$ defined as above. $\Reff$ is determined by the peak of a quadratic fit to $K_{d}$ and $\Lpar$ by the distance between points where the sign of $K_{d}$ reverses \cite{Roberg-Clark2022} within a drift well of ``bad'' curvature, while $a|\bnabla \alpha|$ is evaluated at the center of the fitted drift well, effectively approximating it as a constant. In the small-$b$ limit, $F(b)$ is dominated by the constant term $2.84$, close to the value of $2.66$ in Eqn.~\ref{eqn:abscritgrad}, and as found other works \citep{Romanelli1989,Jenko2001,Roberg-Clark2022} for the case $\tau=1$. In the large $b$ limit, we find, ignoring the small constant $\simeq 0.04$, that the formula effectively predicts $a/L_{T,\text{crit}} \sim a^{2}|\bnabla \alpha|/\Lpar$. Perhaps unsurprisingly, then, the threshold for toroidal modes in this regime is dominated by both the gradient of the bi-normal coordinate (linked to expansion of surfaces as well as local magnetic shear \citep{Roberg-Clark2022a}) and the parallel connection length, which can be reduced by simply increasing the number of field periods in a configuration. Indeed, an experimental realization of this strategy can already be seen in the $10$ field-period LHD heliotron, whose favorable ITG turbulence properties relative to W7-X have been demonstrated \citep{Warmer2021a}. More generally, equations (\ref{eqn:abscritgrad}) and (\ref{eqn:crit}) can now be used to rapidly estimate the absolute and toroidal ITG thresholds on a given magnetic field line.

\begin{figure}
    \centering
    \includegraphics[scale=0.25]{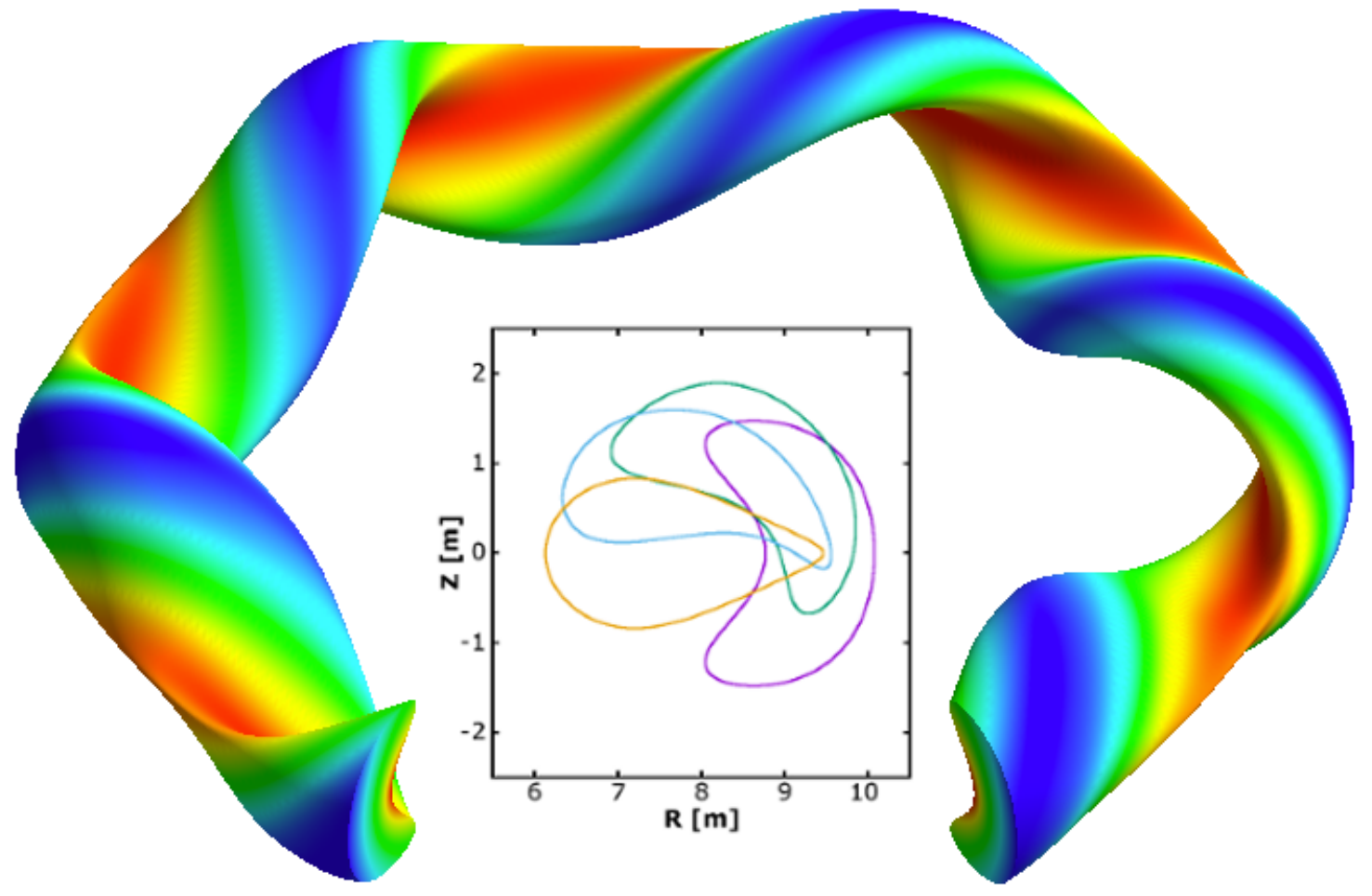}
    \caption{Boundary surface of QSTK (one field period removed), showing contours of $B$ in color. Inset: cuts at constant toroidal angle of the boundary surface in the cylindrical (R,Z) plane.}
    \label{fig:STKBound}
\end{figure}

\textit{Optimization results.}-- We use the SIMSOPT software framework \citep{Landreman2021} to generate two QHS vacuum stellarator configurations. Each stellarator magnetic field is described by a boundary surface given in the Fourier representation $R(\vartheta,\phi)=\sum_{m,n} R_{m,n} \cos(m\vartheta- n_{fp} n\phi), Z(\vartheta,\phi)=\sum_{m,n} Z_{m,n} \sin(m\vartheta- n_{fp} n\phi)$. Optimization proceeds by treating the above-mentioned Fourier coefficients as parameters and varying them in a series of steps in order to find a least-squares minimization of the specified objective function $f$, increasing the number of boundary surface modes with each step. Both optimizations used the ``warm start'' configuration from SIMSOPT with approximate QHS and $n_{fp}=4$. Global zero-$\beta$ equilibria are constructed at each iteration by running the VMEC \citep{Hirshman1983a} code, which solves the MHD equations using an energy-minimizing principle, setting the angular resolution to be $M_{pol}=N_{tor}=7$. 

For the first result, which we call ``HSK'', $f = f_{QS} + (A-4.10)^{2} + f_{\text{abs}}$, where $f_{QS}$ is the quasisymmetry residual defined in \citep{Landreman2022} for QHS with $n_{fp}=4$ at the surfaces $(r/a)^{2} = [0.1,0.2,0.3,0.4,0.5]$, $f_{\text{abs}} = (a/L_{\text{T,crit,abs}} - 2.00)^{2}$ is the critical gradient evaluated at the flux tube $[(r/a)^{2}=0.5,\alpha=0]$, and $f_{A}=(A-4.10)^{2}$ is the aspect ratio target with $A=R/a$ the aspect ratio output by VMEC. The boundary modes varied for the three optimization steps went up to $m_{pol}=n_{tor}=[3,5,6]$. Linear flux tube gyrokinetic simulations with GENE \citep{Jenko2000a} reveal that HSK has the largest critical gradient of any stellarator that we know of, $a/L_{\text{T,crit,abs}}=1.75$ [fig. (\ref{fig:HSK})], as well as relatively low nonlinear ion heat fluxes above that threshold. Further details of HSK and the nonlinear simulations are presented in \citep{Roberg-Clark2022a}. The caveat is that the large ``bad'' curvature of destabilizing sign for HSK (a small $\Reff$ linked to enhancement of $|\bnabla \alpha|)$ produces a vacuum magnetic hill, rendering it Mercier unstable \citep{Mercier1962,Landreman2020} at all values of $\beta$ tested.

In the second optimization we choose $n_{fp}=6$ and target the toroidal ITG threshold in the hopes of reducing turbulent transport while preserving MHD stability. The objective function is
\begin{equation} \label{eqn:objfunc}
    f = f_{QS} + f_{A} + f_{\text{crit}} + f_{\text{well}} + f_{\iota}
\end{equation}
where $f_{A}=[\Theta(A-7.50)]^{2}$ is the aspect ratio penalty, $\Theta(x)$ is defined to be $x H(x)$ with $H(x)$ the Heaviside step function, $f_{QS}$ is again the quasisymmetry residual but with $n_{fp}=6$, and $f_{\text{crit}}=\sum_{\alpha_{j}} [\Theta \left(3.00 - a/L_{T,\text{crit}}(\alpha_{j}) \right)]^{2}$ [eq. (\ref{eqn:crit})] is taken at the surface $r/a=0.5$ and summed over the field lines $\alpha=[0,\pi/8,\pi/4]$, with each field line extending for $8$ poloidal turns, in order to sample the surface. The vacuum magnetic well penalty is $f_{well}=\sum_{r_{k}}[\Theta \left(1 + V''(r_{k})/0.001 \right)]^{2}$ with the surfaces $(r/a)^{2}=[0,0.1,...,0.9]$ targeted and $V''(r)$ the second derivative of the flux surface volume with radius. We calculate the residual $f_{QS}$ on the surfaces $(r/a)^{2}=[0.1,0.2,...,0.9]$, and target the axis and boundary iota via $f_{\iota}=[\iota(r=0)-1.6]^{2} + [\iota(r=a)-1.7]^{2}$, with $\iota$ the rotational transform. The ``warm start'' file was first optimized for increased $ |\bnabla \alpha |$ on the outboard side (see e.g. \cite{Roberg-Clark2022a,Stroteich2022}), and increasing $n_{\text{fp}}$ to 6. The final optimization proceeded in two steps, with the boundary Fourier coefficients varied up to $m_{pol}=n_{tor}=[3,4]$. 

\begin{figure}
    \centering
    \includegraphics[scale=0.6]{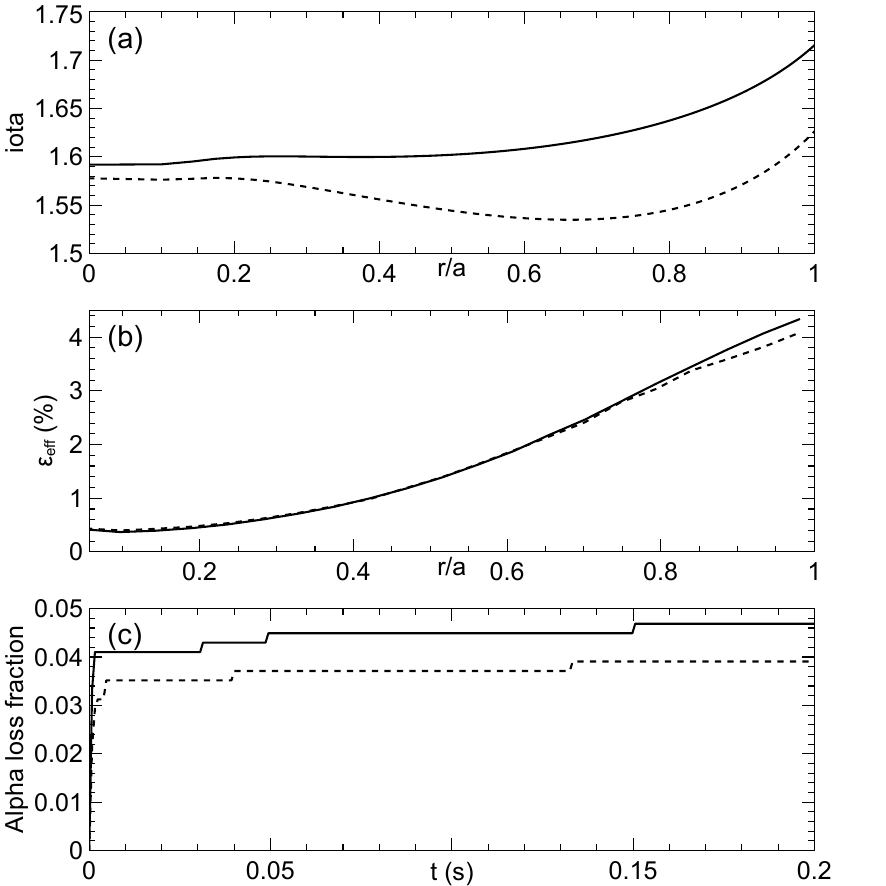} 
    \caption{Properties of QSTK with no coils, in vacuum (solid curves) and at $\beta=1.65\%$ with bootstrap current included (dashed). (a) Rotational transform profile. (b) Neoclassical transport coefficient $\eps_{\text{eff}}$ as a function of radius. (c) Collisionless alpha particle losses at $r/a=0.50$.}
    \label{fig:STKprops}
\end{figure}

The boundary surface for ``QSTK" (Quasi-Symmetric Turbulence Konzept) is shown in Fig. \ref{fig:STKBound}. QSTK has an aspect ratio of 7.5, a volume-averaged magnetic well ($0.7\%$), large rotational transform $> 1.6$, a neoclassical transport coefficient $\eps_{\text{eff}}<1\%$ \citep{Nemov1999} up to roughly half radius, unusually expanded flux surfaces, and $(\simeq 5\%)$ alpha particle losses  $(\simeq 5\%)$ for particles initialized at $(r/a)=0.50$ when QSTK is rescaled to an ARIES-CS-equivalent \citep{Mau2008} minor radius and volume averaged magnetic field strength, using the NEAT code \citep{Jorge2022,Albert2020} [Fig. \ref{fig:STKprops} (c)]. Increased neoclassical transport at the edge [reaching $\eps_{\text{eff}}=4.5\%$, Fig. \ref{fig:STKprops}(b)] may in fact be beneficial, as it can prevent a particle transport barrier from forming, which might otherwise hinder plasma refueling in an experimental scenario \citep{Maasberg1999}. All flux tubes evaluated for QSTK, using the model equation (\ref{eqn:crit}), are predicted to have $a/L_{T,crit} \geq 3.0$.

\begin{figure}
    \centering
    \includegraphics[scale=.30]{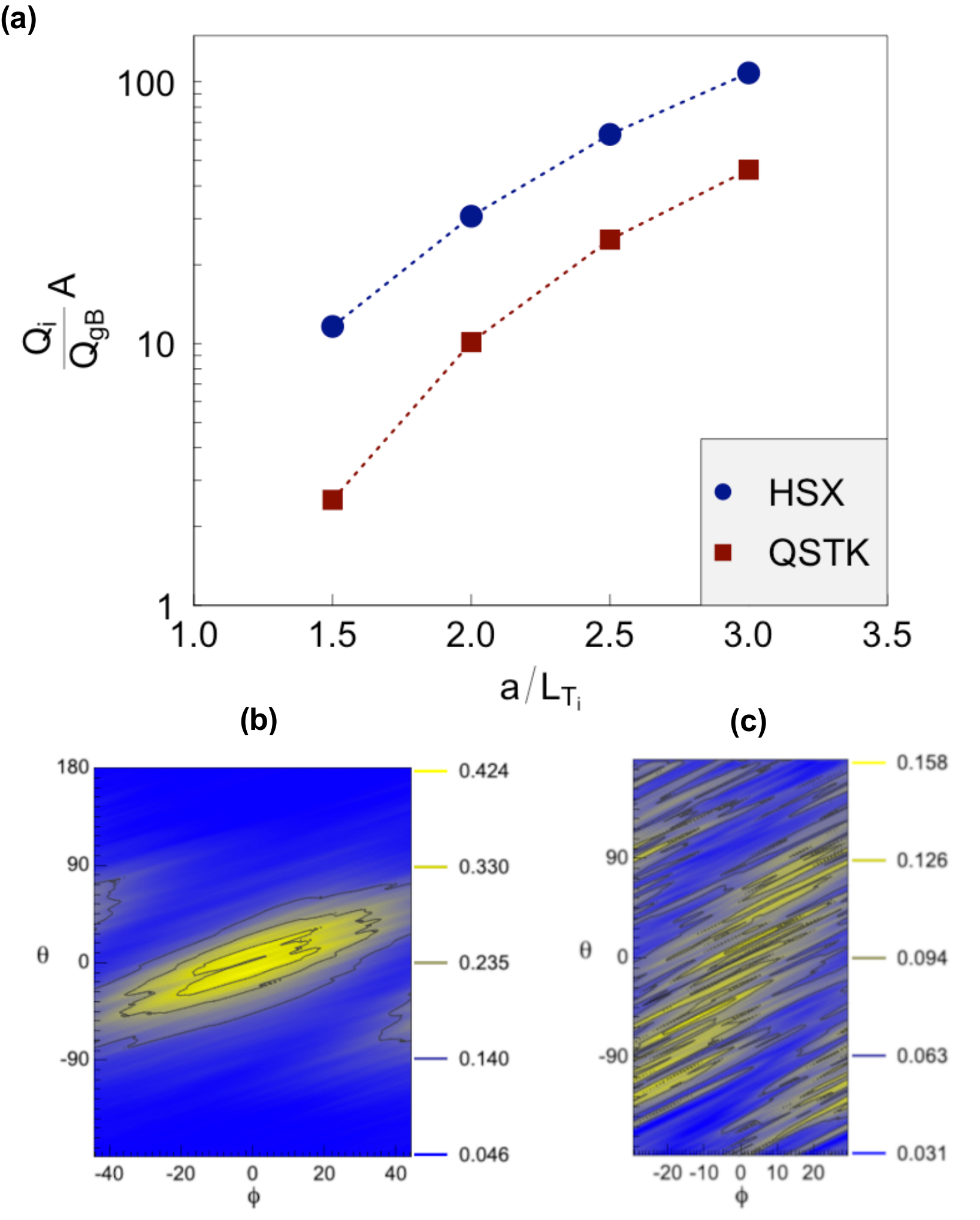} 
    \caption{Full-surface nonlinear gyrokinetic simulations of ITG turbulence comparing HSX to QSTK. (a) Log plot of ion heat flux multiplied by the respective aspect ratio of each configuration. (b) ITG density fluctuations $\tilde{n}/n_{0}$ in one field period of HSX with $a/L_{T}=2$, plotted in Boozer toroidal ($\phi$) versus poloidal ($\theta$) angles. (c) Same as in (b) but for QSTK.}
    \label{fig:ITG}
\end{figure}

\textit{ITG turbulence.}-- To evaluate the performance of QSTK in vacuum with regard to ITG turbulence we run full-surface nonlinear electrostatic gyrokinetic simulations using the GENE code \citep{Jenko2000a,Xanthopoulos2016a} in comparison with the HSX stellarator. We assume adiabatic electrons, zero density gradient, $T_{e}=T$, and temperature gradients $a/L_{T}=[1.5,2.0,2.5,3.0]$ at half radius. In Figs. \ref{fig:ITG} (a)-(b) we plot the ion heat fluxes in gyro-Bohm units times A for each configuration, to adjust for the dependence of energy confinement time on aspect ratio implied by the gyro-Bohm scaling of heat fluxes (a factor of $4/3$ in favor of QSTK). We find that the adjusted heat flux is significantly reduced (by a factor 2-5) in QSTK compared to HSX for the range of gradients studied, demonstrating the success of the optimization strategy. The density fluctuations in QSTK [fig. \ref{fig:ITG}(d)] are relatively weak, and also less localized on the surface, compared to HSX (and most optimized stellarators, e.g. W7-X \citep{Xanthopoulos2020,Wilms2023}), where such fluctuations lie within a strip near the outboard midplane [fig. \ref{fig:ITG}(c)], owing to the more pronounced ``bean-shaped" plane. In contrast, the optimization for QSTK has altered the bean-shaped plane, expanding the surfaces in regions of bad curvature where the toroidal ITG mode resides. We also find $\Lpar \simeq 6 a$ in QSTK versus $ 12 a$ in HSK, suggesting the shortened parallel connection length plays a significant role in the increased value of $b_{min,\text{QSTK}} \sim 3 b_{min,\text{HSX}}$ predicted for QSTK by the model [Eqn. (\ref{eqn:crit})].


\begin{figure}
    \centering
    \includegraphics[scale=.55]{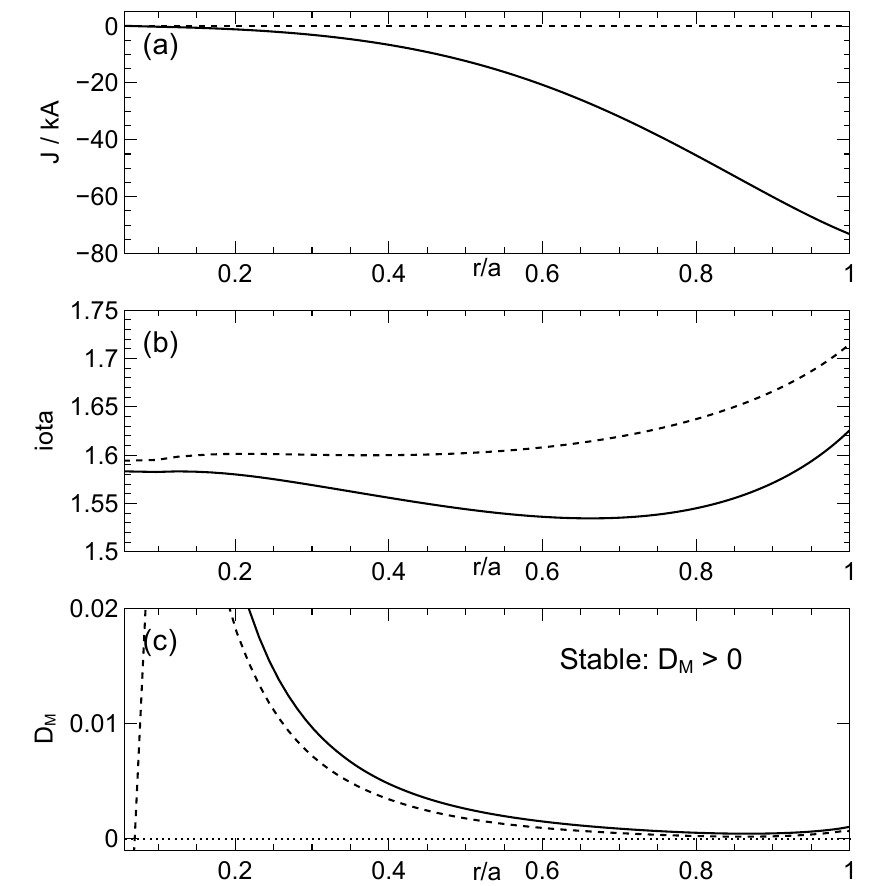} 
    \caption{Bootstrap current and MHD stability of QSTK with an applied pressure profile (see text) at volume-averaged $\beta=1.65\%$. The dashed curves correspond to the case without boostrap current and the solid curves to the case with current. (a) Enclosed toroidal bootstrap current. (b) Rotational transform profile. (c) Mercier stability criterion with a dotted line at $D_{M}=0$.}
    \label{fig:MHD}
\end{figure}

\textit{MHD stability and coils.}-- The QSTK configuration, as a result of the $f_{\text{well}}$ objective in the optimization [eq. \ref{eqn:objfunc})], possesses a vacuum magnetic well and satisfies $V^{\prime \prime}<0$ at all radial locations. An artificial, nearly linear (in $r^{2}$) pressure profile $T=T_{e}=1 \: \text{keV} * [1 - (r/a)^{2}], n_{i}=n_{e}=4.4*10^{20}\: m^{-3}*[1-(r/a)^{10}]$, corresponding to a volume-averaged $\beta=1.65\%$ (with volume-averaged $B\simeq 1 \: T$), is applied to the configuration. The resulting bootstrap current profile calculated with DKES \citep{Hirshman1986,VanRij1989,Beidler2021} amounts to an integrated current of roughly $72 \: \text{kA}$ [fig. \ref{fig:MHD}(a)]. Both the vacuum and bootstrap configurations produce rotational transform profiles that avoid crossing the resonance $m/n = 6/4= 3/2$ [Figs. \ref{fig:STKprops}(a), \ref{fig:MHD}(b)]. The method of \cite{Landreman2022b}, which relies on the isomorphism between quasisymmetry and axisymmetry, was found to produce a similar bootstrap current profile for QSTK, although the results were not in as good agreement as in cases with ``precise" quasisymmetry \citep{Landreman2022}. Alpha losses are slightly reduced to roughly $4\%$ at half radius for the case with bootstrap current [Fig. \ref{fig:STKprops}(c)]. Evaluations \citep{Nuehrenberg1987} of the Mercier criterion indicate that, for the configuration with pressure and bootstrap current included, the Mercier criterion  is satisfied [fig. \ref{fig:MHD}(b)] and the bootstrap current produces a stabilizing up-shift in $D_{M}$ \citep{Ku2011}. Certain radii become Mercier unstable for larger values of $\beta$. We also apply the coil optimization features of SIMSOPT to produce the magnetic field of QSTK, finding that the relative maximum field error can be reduced to $5.1\%$ and the relative mean error to $1.1\%$ with four unique coils (48 coils in total) while penalizing coil length. The initial coil and MHD studies show that QSTK has potential for finite-$\beta$ (reactor-relevant) operation scenarios.

\begin{figure}
    \centering
    \includegraphics[scale=.30]{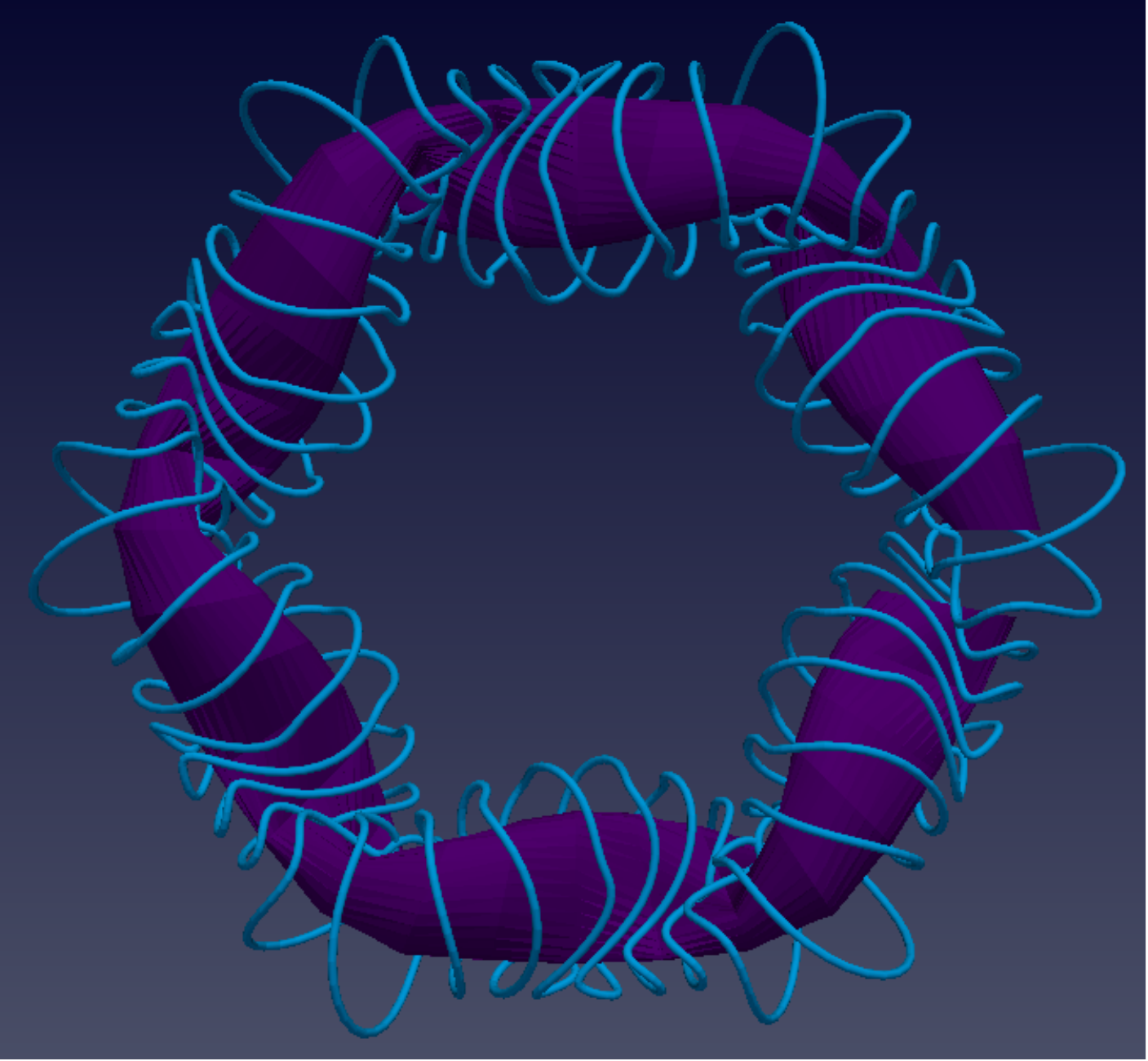} 
    \caption{The boundary surface of QSTK (part of one field period removed) surrounded by electromagnetic coils.}
    \label{fig:coils}
\end{figure}

\textit{Discussion and conclusions.}-- Other microturbulence, such as trapped electron modes and electron temperature gradient (ETG) turbulence, remain to be studied in QSTK. ETG turbulence is likely to benefit from the ITG optimization for QSTK both linearly (from the isomorphism with ITG modes \citep{Jenko2001}) and nonlinearly (from the short connection length \citep{Plunk2019}) with regard to ETG losses. At high $\beta$ values, turbulence is expected to transition from ITG to kinetic ballooning mode turbulence \citep{Aleynikova2018,Pueschel2008a}. This physics is delegated to a separate publication, in the framework of a QSTK reactor study. Despite these open challenges, the present work demonstrates the possibility of modifying the current design of magnetic flux surfaces,
toward a drastic suppression of turbulence in the parameter range usually encountered in modern stellarator experiments. Our method integrates salient physics properties, such as good MHD stability and particle confinement, low neoclassical transport, and bootstrap current, together with the feasibility of modular coils. The QSTK configuration introduced here is thus a contender for a future compact fusion reactor based on the stellarator concept. 



The authors thank P. Helander, A. Zocco, M. Landreman, and C. D. Beidler for helpful conversations. We thank J.F. Lobsien for help with initial coil optimization. This work was supported by a grant from the Simons Foundation (No. 560651, G. T. R.-C.). Computing resources at the RZG (Germany) and the Marconi HPC (Italy) were used to perform the simulations.  This work has been carried out within the framework of the EUROfusion Consortium, funded by the European Union via the Euratom Research and Training Programme (Grant Agreement No 101052200 — EUROfusion). Views and opinions expressed are however those of the author(s) only and do not necessarily reflect those of the European Union or the European Commission. Neither the European Union nor the European Commission can be held responsible for them.

\bibliography{library}


\end{document}